\documentclass[reprint,amsmath,amssymb,pra,longbibliography]{revtex4-1}

\usepackage{graphicx}
\usepackage{subcaption}

\newcommand{\vare}{\varepsilon}
\newcommand{\abs}[1]{\left| #1 \right|}
\newcommand{\brc}[1]{\left[ #1 \right]}

\begin{document}

\preprint{APS/123-QED}

\title{Indirect link between resonant and guided modes 
  on uniform and periodic slabs}

\author{Amgad Abdrabou}
 \email{Corresponding author: mabdrabou2-c@my.cityu.edu.hk}
 \affiliation{Department of Mathematics, City University of Hong Kong,
   Kowloon, Hong Kong, China}
\author{Ya Yan Lu}%
\affiliation{Department of Mathematics, City University of Hong Kong, Kowloon, Hong Kong, China}%
 
\date{\today}

\begin{abstract}
A uniform or periodic dielectric slab can serve as an optical
waveguide for which guided modes are important, and it can also be
used as a diffraction structure for which resonant modes with complex
frequencies are relevant. Guided modes are normally studied below the
lightline where they exist continuously and emerge from points on the
lightline, but isolated guided modes may exist above the lightline and
they are the so-called bound states in the continuum. Resonant modes
are usually studied above the 
lightline (defined using the real part of the complex frequency), but
they are not connected to the guided modes on the lightline. 
In this work, through analytic and numerical calculations for uniform
and periodic slabs, we establish an indirect link between the resonant and
guided modes. It is shown that as the (Bloch) wavenumber is increased,
a resonant mode continues its existence below the lightline, until it
reaches its end at an Exceptional Point (EP) where a pair of improper
modes emerge, and one branch of improper modes eventually approaches
the lightline at the starting point of a guided mode. Leaky modes with a
real frequency  and a complex (Bloch) wavenumber (propagation
constant) are also related to the improper modes. They emerge at EPs in
eigenvalue formulations where the frequency is regarded as a parameter. Our study is based on a non-Hermitian eigenvalue formulation that
includes resonant, improper and leaky modes, 
and provides a complete 
picture for different kinds of eigenmodes on uniform and periodic slabs.
 \begin{description}
 \item[PACS numbers]
 42.65.Hw,42.25.Fx,42.79.Dj
 \end{description}
\end{abstract}

\maketitle


\section{Introduction}\label{S1}

Eigenvalue problems for photonic structures such as optical
waveguides, resonant cavities and periodic media are of fundamental
importance \cite{Marcuse,MITPhC}. Although the research topic has a
long history, new and interesting wave phenomena related to eigenvalue
problems are continuously being discovered.  Recently, bound states in
the continuum (BICs) \cite{Von29}, which are trapped or guided modes
with their frequencies within the radiation continuum, have attracted
much attention and found a number of significant applications in
optics \cite{Hsu13,BIC,kodi17}.  Exceptional points (EPs)
\cite{Kato,Heiss}, which are spectral degeneracies for non-Hermitian
eigenvalue problems with coalescing eigenvalues and eigenfunctions,
are being intensively investigated in optics
\cite{BoZhen,Arslan,Zhou2018,Amgad1}, and may have valuable
applications in lasing and sensing \cite{Chen2017,Hodaei}. EPs are
also the key to many unusual and counter-intuitive wave phenomena in
${\cal PT}$-symmetric and related structures
\cite{PTsym,Feng,EPLas2,Goldzak}.  For dielectric structures that are
unbounded in one or two spatial directions, 
the most important eigenvalue problems are those for guided and
resonant modes \cite{Fan}.  
Guided modes  decay exponentially in the media surrounding the
structure, and usually exist below the lightline.  The resonant
modes radiate power to infinity,  have complex frequencies, and are usually 
studied above the lightline. The BICs are isolated guided modes above the 
lightline, and they can also be regarded as special resonant modes with 
infinite quality factors \cite{BIC}. EPs of resonant modes have been found on
various open dielectric structures
\cite{BoZhen,Chen2017,Arslan,Zhou2018,Amgad1}.  This is not
unexpected, since the eigenvalue problems for resonant modes on open
structures are non-Hermitian due to radiation losses. 


For unbounded structures
that are periodic or invariant in one or two spatial directions,
band structures for guided and resonant modes have been calculated by
many authors. Below the lightline, there are continuous branches
of guided modes, and they emerge from starting points on the
lightline. Resonant modes are often calculated above the lightline,
and they seem to approach the lightline near the starting points of
the guided modes. However, careful numerical studies
show that the guided modes and the resonant modes are not connected on 
the lightline. In fact, 
the  resonant  modes cross the lightline 
with a nonzero imaginary part in their complex frequencies.
We believe it is interesting to study the resonant modes below the
lightline, and establish a link between 
the resonant and guided modes. To the best of our knowledge, such a
link has never been established before.  
To find the missing link between these
two kinds of modes, we concentrate on simple uniform or periodic
dielectric slabs in this paper. Our study relies on a mathematical
formulation that includes  
not only the resonant modes, but also improper modes that grow
exponentially in the surrounding media and leaky modes with real
frequencies and complex wavenumbers~\cite{Marcuse}.  Our
analysis 
reveals that as the wavenumber is increased, a typical branch of
resonance modes continues its existence below the lightline until it
reaches an EP where a pair of improper 
modes emerge, and one branch of the improper modes eventually ends up
on the lightline where it connects to the guided modes. 

The rest of this paper is organized as follows. In Sec.~II, we present
some basic properties for eigenvalue problems on one-dimensional (1D)
structures for which a uniform slab is a special case. 
In Sec.~III, we show numerical results for different types of
eigenmodes on a slab with a constant refractive index. Section IV is
devoted to a periodic slab with a piecewise constant refractive index
profile. The paper is concluded by some remarks and discussions in Sec.~V.

\section{Eigenmodes on 1D structures}
\label{S2}

\begin{figure}[htb]
	\centering
	\includegraphics[width=0.8\linewidth]{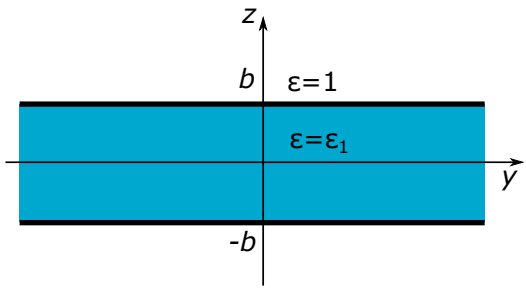}
	\caption{A uniform slab of thickness $h=2b$ and relative
          permittivity  $\vare_1$ surrounded by air.} 
	\label{Fig1}
 \end{figure}

We consider a 1D dielectric wave-guiding structure surrounded by air.
A Cartesian coordinate system is chosen so that the  structure is 
perpendicular to the $z$ axis. The relative permittivity depends only
on $z$ and satisfies $\varepsilon(z)= 1$ for $|z|>b$ and  
$\varepsilon(z) = \varepsilon_1(z)$ for $|z|<b$, where $\varepsilon_1(z) 
\ge 1$,  $\max [ \varepsilon_1(z)] > 1$, and $h=2b$ is the thickness of
the structure. A uniform slab, shown in Fig.~\ref{Fig1} 
below,  is a special case for which $\varepsilon_1(z)$ is a constant.

Assuming the field is invariant in the $x$
 direction, a transverse electric (TE) eigenmode of the structure is a solution with its
electric field component $E_x$ given as
\begin{equation}
\label{timed}
E_x = \mbox{Re} [ \phi(z)\, e^{i(\beta y-\omega t)}],   
\end{equation}
where $\phi(z)$ is the mode profile, $\beta$ is the propagation
constant, $\omega$ is the angular frequency, $k=\omega/c$ is the free
space wavenumber, and $c$ is the speed of light in vacuum. The function
$\phi$ satisfies the following 1D Helmholtz equation 
\begin{equation}
\label{Eq1}
\frac{d^2\phi }{dz^2}+ k^2\vare(z)\phi =  \beta^2\phi,
\end{equation}
for $-\infty < z < \infty$, subject to boundary conditions that
determine the physical nature of the solution.  


Guided modes are solutions for real $\beta$ and real $k$, such that 
$\phi(z)\to 0$ as $z\to \pm \infty$. For $|z| > b$, Eq.~(\ref{Eq1}) is
reduced to $d^2\phi/dz^2 + (k^2-\beta^2) \phi = 0$, and the solutions
are either oscillatory or exponential depending on whether $k^2 -
\beta^2$ is positive or negative. Therefore, guided modes only exist 
when $ k < |\beta|$, that is, below the lightline. 
Moreover, the guided modes satisfy the following boundary condition
\begin{equation}
\label{Eq4}
\frac{d\phi }{dz} = \mp \gamma_0 \phi \quad \mathrm{at}\quad z = \pm
b,  
\end{equation}
where $\gamma_0 = \sqrt{\beta^2-k^2}$. 
Equation (\ref{Eq1}) for $|z| < b$ and Eq.~(\ref{Eq4}) give rise to an
eigenvalue problem of the guided modes on the finite interval
$(-b,b)$ for $z$. Assuming $\beta$ is given, then $k$ or $k^2$ can be regarded as the
eigenvalue. Alternatively, if $k$ is given, then $\beta$ or $\beta^2$
is the eigenvalue. Since $\gamma_0$ depends on the eigenvalue, 
the above formulation on the finite interval of $z$ is nonlinear. 

For a guided mode $\{\beta, k, \phi(z) \}$, the derivative of $k$ with
respect to $\beta$ is given by  
\begin{equation}
  \label{dkdb}
\frac{dk}{d\beta} = \frac{ \beta \int_{-\infty}^\infty
  |\phi(z)|^2 dz}{ k \int_{-\infty}^\infty 
  \epsilon(z) |\phi(z)|^2 dz}.
\end{equation}
The above
can be derived by taking the partial derivative of
Eq.~(\ref{Eq1}) with respect to $\beta$,  multiplying the resulting
equation by $\overline{\phi}$ (the complex conjugate of $\phi$) and
integrating with respect to $z$ on $(-\infty,\infty)$. It is clear that for $\beta > 0$,
$dk/d\beta > 0$, thus, the dispersion curves of guided modes always
have a positive slope. Equation (\ref{dkdb}) can be written  as
\[
\frac{dk}{d\beta} = \frac{\beta}{k} \cdot \frac{ 2 \gamma_0 \int_{-b}^b |\phi(z)|^2
  dz + |\phi(b)|^2 + |\phi(-b)|^2}{ 2 \gamma_0 \int_{-b}^b
  \epsilon(z) |\phi(z)|^2 dz + |\phi(b)|^2 + |\phi(-b)|^2}.
\]
Therefore, as the dispersion curve approaches the lightline for $\beta
> 0$, we have $\beta/k \to 1$, $\gamma_0 \to 0$ and $dk/d\beta \to
1$. In other words,  at the limiting points on the lightline, the dispersion
curves of guided modes are tangential to the lightline. 


For real $\beta$ and real $ k < |\beta|$, Eq.~(\ref{Eq1}) also has
solutions that blow up at infinity, i.e.,  $\phi(z) \to \infty$ as $z
\to \pm \infty$. These solutions have been referred to as improper
modes \cite{Yama,Hanson}, and they satisfy the boundary condition
\begin{equation}
\label{Eq6}
\frac{d\phi }{dz} = \pm \gamma_0 \phi \quad \mathrm{at}\quad 
z = \pm b.  
\end{equation}
The improper modes are unphysical when they are considered on the
infinite interval of $z$, but they have physical meaning on any
bounded interval $(-D,D)$ for $D > b$. If we put proper waveguides
with high 
refractive indices in $z>D$ and $z< -D$, respectively, a supermode of
the multilayered structure may have the same $\beta$ and same $k$ as
the improper mode, and the mode profile of the supermode is then
identical to that of the improper mode for $z \in (-D,D)$.


Resonant modes are solutions  of the Helmholtz equation for real $\beta$ and
complex $k$ (i.e., complex $\omega$) that exhibit outgoing wave behavior
as $z \to \pm \infty$. They radiate power to infinity and decay with
time. Under the time dependence given in Eq.~(\ref{timed}), we have
$\mbox{Im}(\omega) < 0$. 
If we define $\alpha_0 = \sqrt{k^2 - \beta^2}$ based on the standard
complex square root with the real negative axis as the branch cut, then 
$\phi(z) = \phi(b)e^{ i \alpha_0 (z-b)}$ for $z > b$ and 
$\phi(z) = \phi(-b)e^{ -i \alpha_0 (z+b)}$ for $z <-b$. Therefore, the
resonant modes satisfy the following boundary
 conditions 
\begin{equation}
\label{EqRa}
\frac{d\phi }{dz} = \pm i \alpha_0 \phi \quad
\mathrm{at}\quad z = \pm b.
\end{equation}
Multiplying Eq.~(\ref{Eq1}) by $\overline{\phi}$, integrating the
result on $(-b,b)$, and applying the boundary condition (\ref{EqRa}), we obtain 
\begin{multline}
\label{Eq7}
i \alpha_0 \brc{ \abs{\phi(b)}^2+  \abs{\phi(-b)}^2} 
\\=  \int_{-b}^{b}
\abs{\frac{d\phi}{dz}}^2 dz+ \  \int_{-d}^{d}\brc{\beta^2-k^2
  \vare_1(z) }\abs{\phi}^2 dz.
\end{multline}
The imaginary part of above gives 
\begin{equation}
  \label{imagk}
\mbox{Im}(k) = \frac{ - \mbox{Re}(\alpha_0) \brc{ \abs{\phi(b)}^2+
    \abs{\phi(-b)}^2} }
{2 \mbox{Re}(k) \int_{-b}^b \varepsilon_1(z) |\phi(z)|^2 dz}.
\end{equation}
If $k$ is complex, then $\mbox{Re}(\alpha_0) > 0$ due to the
definition of the square root, and Eq.~(\ref{imagk}) gives
$\mbox{Im}(k) < 0$. This is consistent with the requirement that the
resonant modes should decay with time. In addition, we have
$\mbox{Im}(\alpha_0) < 0$, thus $\phi$ blows up as $z \to \pm \infty$. 
The resonant modes are usually
studied above the lightline, that is for $\mbox{Re}(k) > |\beta|$. In
general, as $\mbox{Re}(k) \to \beta$, $\mbox{Im}(k)$ and $\alpha_0$
do not tend to $0$, thus, the resonant modes and guided modes are not
connected on the lightline. In latter sections, we show that the
resonant modes continue their existence below the lightline.  


If $\{k, \phi\}$ is a resonant mode for a real $\beta$, then
$\{\bar{k},\bar{\phi} \}$ is another solution of Eq.~(\ref{Eq1}) for
the same $\beta$. If we denote that solution also by $\{k, \phi \}$, 
then it satisfies 
\begin{equation}
\label{EqRb}
\frac{d\phi }{dz} = \mp i \alpha_0 \phi \quad
\mathrm{at}\quad z = \pm b. 
\end{equation}
For the time dependence given in Eq.~(\ref{timed}), a solution
satisfying Eqs.~(\ref{Eq1}) and (\ref{EqRb}) represents the time
reversal of a resonant mode. The field consists of incoming waves from
$z=\pm \infty$ and its amplitude grows with time. On the other hand,
if we replace the the time dependence by $e^{i \omega t}$, then a 
resonant mode satisfies Eq.~(\ref{EqRb}) and its time reversal
satisfies Eq.~(\ref{EqRa}).  We point out that both the resonant modes and
their time reversals satisfy boundary condition (\ref{Eq6}), with
$\gamma_0 = \sqrt{\beta^2 - k^2}$ defined using the standard complex
square root. This is so, because for a resonant mode,  $\mbox{Im}(k) <
0$, thus $\mbox{Im}(\beta^2 - k^2) > 0$,  $\mbox{Im}(\gamma_0) > 0$,
and $\gamma_0 = i \alpha_0$;  and for the time reversal of the resonant
mode, $\mbox{Im}(k) > 0$, thus $\mbox{Im}(\beta^2 - k^2) < 0$,
$\mbox{Im}(\gamma_0) < 0$, and $\gamma_0 = -i \alpha_0$. Therefore,
the boundary condition (\ref{Eq6}) allows us to unify the improper
modes, the resonant modes and their time reversals. Actually, this is
quite natural, since all these modes blow up as $z \to \pm \infty$. 


For optical waveguides, leaky modes with real frequencies and complex
propagation constants are widely studied \cite{Marcuse}. 
Under the dependence on $t$ and $y$ given in Eq.~(\ref{timed}), the
mode profile satisfies the same Helmholtz equation (\ref{Eq1}). Notice
that a leaky mode with $\mbox{Re}(\beta) > 0$ should have $\mbox{Im}(\beta) > 0$,
so that as it propagates forward in the positive $y$ direction, it radiates power to
infinity (for $z \to \pm \infty$), and its amplitude decreases. If $\{
\beta, \phi\}$ is a leaky mode, then its complex conjugate $\{
\bar{\beta}, \bar{\phi} \}$ represents  a time reversed solution that increases as it
propagates forward by gaining power coming from infinity (i.e., from
$z=\pm \infty$).  Alternatively, $\{ \bar{\beta}, \bar{\phi} \}$ can
also be regarded as a leaky mode with the dependence on $t$ and $y$
switched to $e^{i   (\omega t - \beta y)}$. Once again, the boundary
condition \eqref{Eq6} is applicable to both leaky modes  and their
time reversals. 


In summary, a 1D dielectric wave-guiding structure supports guided and
improper modes for real $\beta$ and real $k$, resonant modes and their
time reversals for real $\beta$ and complex $k$, leaky modes and their
time reversals for real real $k$ and complex $\beta$. Except for the
guided modes, all other modes blow up as $z \to \pm \infty$ and
satisfy boundary condition (\ref{Eq6}).

\section{Uniform slab: numerical results}

In this section, we present numerical results for a uniform slab with
a constant refractive index, that is, $\varepsilon(z) =
\varepsilon_1$, a constant, for $-b<z<b$. Due to reflection symmetry
with respect to the mid-plane of the slab, the eigenmodes are
either even in $z$ or odd in $z$. For the even and odd modes, 
the mode profile can normalized by the condition $\phi(b)=1$, and be
written as  
\[
\phi(z) = \frac{ \cos (\gamma_1 z)}{\cos(\gamma_1 b)} \quad \mbox{or}
\quad 
\phi(z) = \frac{\sin(\gamma_1 z)}{\sin(\gamma_1 b)}
\]
for $|z| < b$, respectively, where  $\gamma_1 = \sqrt{k^2 \vare_1
  -\beta^2}$. 

For the guided modes, we have $\phi(z) = e^{ - \gamma_0 (z-b)}$ 
for $z > b$, then the continuity of $d\phi/dz$ at $z=b$ gives us the
following equation
\begin{equation}
\label{EqG}
e^{2i\gamma_1 b} \mp \frac{i\gamma_0+\gamma_1}{i\gamma_0-\gamma_1}=0,
\end{equation}
where the $+$ and $-$ signs correspond to the even and odd modes,
respectively. The guided modes form discrete branches below the
lightline, and  $m$-th dispersion curve approaches the
lightline ($k=\beta$) at  
\begin{equation}
k = \beta = k_m = \frac{\pi P_m}{d \sqrt{\vare_1-1}}  
\end{equation}
where $P_m = m$ for the even modes and  $P_m = m-1/2$ for the odd
modes, and $m \ge 1$. 

For the improper, resonant and leaky modes, and their time reversals, we
need to use boundary condition (\ref{Eq6}), and thus $\phi(z)  = e^{
  \gamma_0 (z-b)}$ for $z > b$. From the continuity of $d\phi/dz$ at
$z=b$, we obtain 
\begin{equation}\label{EqR}
e^{2i\gamma_1 b} \mp \frac{\gamma_0+i\gamma_1}{\gamma_0-i\gamma_1}=0.
\end{equation}
Again, the $+$ and $-$ signs correspond to the even and odd modes,
respectively.  

\begin{figure}[htbp]
	\captionsetup[subfigure]{justification = centering}
	\centering
   \begin{subfigure}[b]{0.95\linewidth}
   \includegraphics[width=\textwidth]{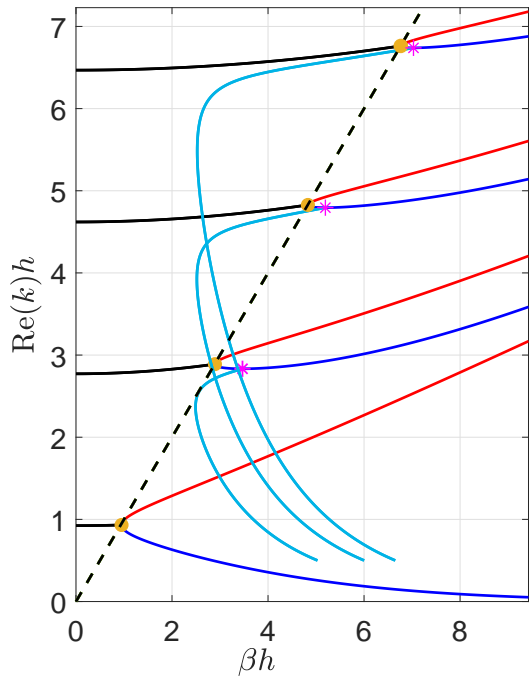}
		\caption{}	
		\label{Fig2a}	
	\end{subfigure}
	\begin{subfigure}[b]{0.95\linewidth}
		\includegraphics[width=\textwidth]{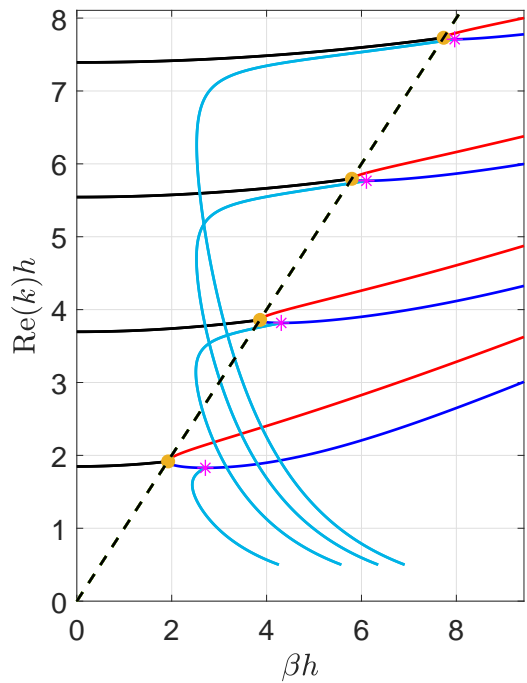}
		\caption{}
	\label{Fig2b}
	\end{subfigure}
	\caption{Dispersion curves of  (a) odd and (b) even modes for
		$1\leq m \leq 4$, showing guided (red), resonant (black),
		improper (blue) and leaky (cyan) modes.}\label{Fig2}	
\end{figure}

To find the dispersion relations of various modes, we
only have to solve Eqs.~(\ref{EqG}) and (\ref{EqR}). 
As a numerical example, we consider a slab with $\vare_1=11.56$. In Figs.~\ref{Fig2a} and~\ref{Fig2b},
we show the dispersion curves of the first four odd and even modes,
respectively.  In particular, we show $\mbox{Re}(k)h$ for the resonant
modes and $\mbox{Re}(\beta)h$ for the leaky modes. The guided,
resonant, leaky and improper modes are shown in red, black, cyan and
blue, respectively.  Notice that there is no odd leaky mode for $m=1$. 

\begin{figure}[h!]
	\centering 
	\includegraphics[width=\linewidth]{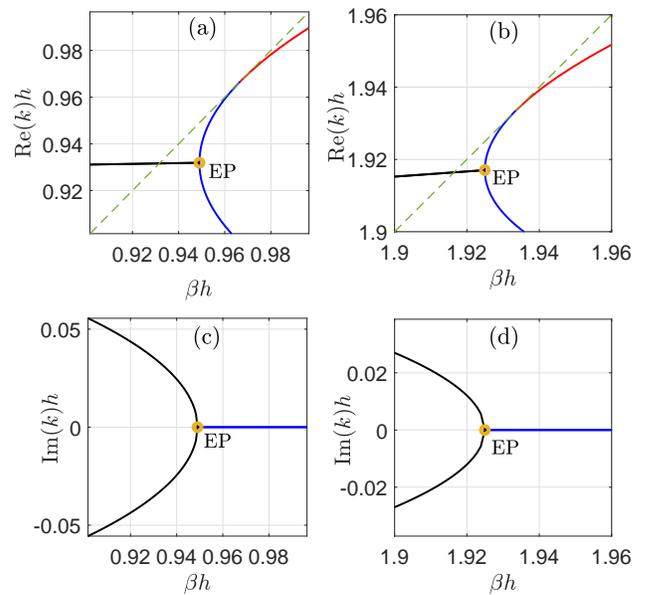}
	\caption{Real parts (Figs.~\ref{Fig3}a and~\ref{Fig3}b) and imaginary parts (Figs.~\ref{Fig3}c and~\ref{Fig3}d) of $kh$ versus $\beta h$ for the odd and even modes with $m=1$, respectively.}
	\label{Fig3}
\end{figure}

From Fig.~\ref{Fig2}, we observe that the resonant modes approach the
lightline near the starting points of the
guided modes, but actually these two kind of modes are not connected
on the lightline. In Fig.~\ref{Fig3}, 
we show the first odd and even modes near the lightline. The red,
black and blue curves correspond to guided, resonant and improper
modes, respectively. The guided modes converge to the lightline
tangentially, confirming the theoretical result given in Sec.~II. 
For the resonant modes (and their time reversals), we show
$\mbox{Re}(kh)$ in Figs.~\ref{Fig3}a and \ref{Fig3}b, and $\mbox{Im}(kh)$ in
Figs.~\ref{Fig3}c and \ref{Fig3}d. Figures~\ref{Fig3}a and \ref{Fig3}c show the first odd mode, and
Figs.~\ref{Fig3}b and \ref{Fig3}d correspond to the first even mode.  Notice that the
resonant mode and its time reversal exist below 
the lightline until they reach an EP, beyond which a pair of improper
modes emerge.  More precisely, as $\beta$ increases through the EP, a
complex conjugate  pair of $k$ (for a resonant mode and its time reversal) is turned to 
two different real values of $k$ corresponding to two improper modes. 
In addition, $\mbox{Im}(k)$ and $\mbox{Re}(k)$ exhibit a one-sided 
square root splitting before and after the EP, respectively. 
When $\beta$ is further increased, one branch of the improper modes
converge to the lightline at exactly the same point where a branch of guided
modes emerges. The higher order modes (with $m > 1$) have exactly the 
same behavior. 


The EPs separating resonant and improper modes can be easily
calculated, since they correspond to $dk/d\beta  = \infty$. From
Eq.~(\ref{EqR}) and its derivative, we obtain the following system
\begin{equation}\label{Eq15}
\begin{aligned}
& \gamma_1+\vare_1 \gamma_0\cot(\gamma_1 b)- \vare_1 b
\gamma_1\gamma_0 \csc^2(\gamma_1 b)=0,\\ 
&\gamma_1\cot(\gamma_1 b)-\gamma_0 =0
\end{aligned}
\end{equation}
for EPs of odd modes, and 
\begin{equation}\label{Eq17}
\begin{aligned}
& \gamma_1 -\vare_1 \gamma_0 \tan(\gamma_1 b)- \vare_1  b
\gamma_1\gamma_0 \sec^2(\gamma_1 b)=0,\\ 
&\gamma_1\tan(\gamma_1 b)+\gamma_0 =0
\end{aligned}
\end{equation}
for EPs  of even modes. The above systems can be easily solved. The
results are listed in the following table. 
\begin{table}[htb]
  \centering
  \begin{tabular}{c|c|c|c} \hline
    Parity & $m$ & $\beta h$ & $k h$ \\ \hline
    odd & $1$& $0.9490247327$ & $0.9319093681$ \\ \hline
    odd & $2$& $2.8945994396$ & $2.8893869146$ \\ \hline
    odd & $3$& $4.8303964776$ & $4.8272891351$ \\ \hline
    odd & $4$& $6.7648851107$ & $6.7626694968$ \\ \hline
    even &$1$& $1.9249524220$ & $1.9170319451$ \\ \hline
    even &$2$& $3.8627853763$ & $3.8588932776$ \\ \hline
    even &$3$& $5.7977220843$ & $5.7951354860$ \\ \hline
    even &$4$& $7.7319467271$ & $7.7300088986$ \\ \hline
  \end{tabular}
  \caption{Exceptional points in the $\beta$-$k$ plane separating resonant
    modes and improper modes.} 
  \label{tab1}
\end{table}

\begin{figure}[htb]
	\centering 
	\includegraphics[width=\linewidth]{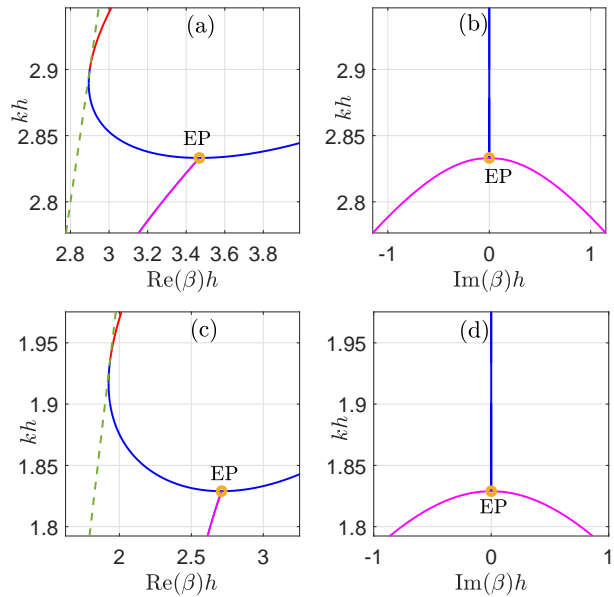}
	\caption{Normalized free-space wavenumber $kh$ versus $\mathrm{Re}(\beta)h$ (Figs.~\ref{Fig4}a and~\ref{Fig4}c) and $\mathrm{Im}(\beta)h$ (Figs.~\ref{Fig4}b and~\ref{Fig4}d) for the leaky odd ($m=2$) and even ($m=1$) modes, respectively. }
	\label{Fig4}
\end{figure}

Leaky modes with real $k$ and complex $\beta$ are also shown in
Fig.~\ref{Fig2}. It is known that the leaky modes are connected to the
improper modes  \cite{Yama,Hanson}. In fact, leaky modes 
emerge from local minima of the dispersion curves of the improper 
modes. For the uniform slab considered in this section, 
there is no leaky mode associated with the lowest odd improper mode
($m=1$), since for that mode, $k$ decreases monotonically as $\beta$
is increased.  In  Figs.~\ref{Fig4}a and~\ref{Fig4}c, 
we show the dispersion curves,  $kh$ versus $\mbox{Re}(\beta)h$, for the
second odd mode ($m=2$) and the 
first even mode ($m=1$). The imaginary parts of $\beta h$ are
shown in Figs.~\ref{Fig4}b and \ref{Fig4}d. The red,
blue and magenta curves correspond to the guided, improper and leaky
modes, respectively. It is clear that as $k$ is decreased, two real
$\beta$ (for two improper modes) are turned to a complex conjugate pair of $\beta$
through an EP. The  complex conjugate pair of $\beta$ represents a
leaky mode and its time reversal.

The EPs separating the leaky and improper modes can also be
easily calculated, since they satisfy the necessary condition $dk /
d\beta = 0$. From Eq.~(\ref{EqR}) and its derivative, we obtain the
following system
\begin{equation}\label{Eq16}
\begin{aligned}
& \gamma_1+ \gamma_0 \cot(\gamma_1 b)- b \gamma_1\gamma_0
\csc^2(\gamma_1\, b)=0,\\ 
&\gamma_1\cot(\gamma_1\, b)-\gamma_0 =0
\end{aligned}
\end{equation}
for EPs of the odd modes, and 
\begin{equation}\label{Eq18}
\begin{aligned}
& \gamma_1  - \gamma_0 \tan(\gamma_1\, b)- b \gamma_1\, \gamma_0
\sec^2(\gamma_1\, b)=0,\\ 
&\gamma_1\tan(\gamma_1\, b)+\gamma_0 =0
\end{aligned}
\end{equation}
for EPs of the even modes. 
 
\section{Periodic slab\label{S3}}

\begin{figure}[h!]
	\centering
	\includegraphics[width=0.8\linewidth]{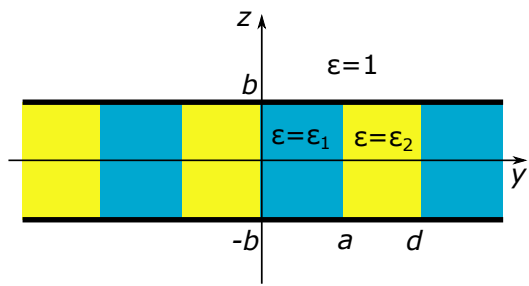}
	\caption{A dielectric slab that is invariant in $x$ and
          periodic in $y$ with period $d$. Each
		period consists of two uniform segments.}
	\label{Fig8}
\end{figure}

In this section, we consider a simple two-dimensional (2D) dielectric
structure, that is, a periodic slab with two uniform segments, and show that
the resonant and guided modes of the periodic slab are also linked
under the lightline through EPs and 
improper modes. A schematic of the periodic slab is shown in
Fig.~\ref{Fig8}, 
where $z$ is perpendicular to the slab, the $xy$ plane is the
mid-plane of the slab, $h=2b$ is the slab thickness, $d$ is the period in the $y$
direction, $a$ and $d-a$ are the widths of the two uniform segments
with dielectric constants $\varepsilon_1$ and
$\varepsilon_2$, respectively. The slab is invariant in the $x$
direction and surrounded by air (for $z> b$ and $z< -b$).  

Similar to Sec.~II, we give definitions and boundary conditions
for various modes on the periodic slab. 
For the $E$-polarization, the $x$ component of the electric field is 
assumed to be $E_x = \mbox{Re} ( u\, e^{ - i \omega t})$, where 
$u=u(y,z)$ satisfies the following 2D Helmholtz equation
\begin{equation}
\label{Eq16p}
\partial_y^2 u +\partial_z^2 u +k^2 \vare(y,z)u = 0,
\end{equation}
and $\varepsilon$ is the dielectric function. 
A Bloch mode of the periodic slab is a solution of Eq.~\eqref{Eq16p}
given in the form 
\begin{equation}
\label{Eq17p}
u(y,z) = \phi(y,z)\,e^{i\beta y},
\end{equation}
where $\beta$ is  the Bloch wavenumber and $\phi(y,z)$ is periodic in
$y$ with period $d$. Thus, $u$ is quasi-periodic in $y$ satisfying 
\begin{equation}
  \label{qperiodic}
u(y+d,z) = e^{ i \beta d} u(y,z).
\end{equation}

On a lossless dielectric slab, a guided Bloch mode has a real $\beta$ and
a real $k$ (i.e., a real frequency), and decays exponentially away from
the slab (i.e., as $z\to \pm \infty$). Due to the
periodicity in $y$, it is only necessary to consider $\beta$
in the interval $[-\pi/d, \pi/d]$. Below the lightline (i.e., for $k < |\beta|$), guided modes
exist continuously with respect to $\beta$ or $k$. For $|z| > b$, the
wave field of a guided mode (below the lightline) can be expanded in
evanescent plane waves as
\begin{equation}
  \label{uexp}
u(y,z) = \sum_{m=-\infty}^\infty \hat{u}_m^{\pm} e^{ i \beta_m y \mp \gamma_m z}, \quad \pm z > b,  
\end{equation}
where $\hat{u}_m^\pm$ are expansion coefficients, and 
\begin{equation}
  \label{dfbtgm}
\beta_m = \beta + 2\pi m/d,   \quad  \gamma_m = \sqrt{ \beta_m^2 -
  k^2}. 
\end{equation}
Notice that $\beta_0 = \beta$, all $\gamma_m$ are positive, $\gamma_0$
is identical to that defined in Sec.~II, but $\gamma_1$ is different from that of Sec.~III. If we
define a linear operator $\Lambda=\Lambda(\beta,k)$, such that 
\begin{equation}
  \label{dfLam}
\Lambda e^{ i \beta_m y} = - \gamma_m e^{ i \beta_m y}, \quad m=0,
\pm 1, \pm 2, ...  
\end{equation}
then the guide mode satisfies the following boundary condition
\begin{equation}
  \label{bcguided}
\partial_z u = \pm \Lambda u, \quad z = \pm b.  
\end{equation}

A Bloch resonant mode on any 2D open periodic structure is a solution that
radiates power to infinity and decays with time.  It is given in
Eq.~(\ref{Eq17p}) for a real $\beta$ and a complex $k$ (or $\omega$)
with $\mbox{Im}(k) < 0$. Resonant modes are usually studied above the
lightline, i.e., for $\mbox{Re}(k) > |\beta|$. For $|z|>b$, the wave
field of a resonant mode can also be expanded, but one or more terms in
Eq~(\ref{uexp}) must be replaced by outgoing waves with growing
amplitudes. For simplicity, we concentrate on resonant modes 
with only one outgoing wave component. In that case, the wave field
can be written as 
\begin{eqnarray}
\nonumber 
u(y,z) &=& \hat{u}_0^\pm e^{ i \beta_0 y \pm  \gamma_0 z}  \\
  \label{uexp0}
& +&  \sum_{m \ne   0} \hat{u}_m^{\pm} e^{ i \beta_m y \mp \gamma_m  z}, \quad \pm z > b.
\end{eqnarray}
Notice that the sign for $\gamma_0 z$ is switched. In the above, 
$\beta_m$ and $\gamma_m$ are still given in Eq.~(\ref{dfbtgm}), but 
all $\gamma_m$ are complex and they are defined using the 
standard complex square root with a branch cut on the negative real
axis. Since 
$k$ has a negative imaginary part, $\beta^2 - k^2$ is in the upper
complex half-plane,  thus $\gamma_0$ is in the first quadrant. Since
$\mbox{Im}(\gamma_0) > 0$, the term $e^{  \gamma_0 z}$ represents a
plane wave propagating towards $z=+\infty$. 
Since $\mbox{Re}(\gamma_0)$ is also positive, the amplitude of that
plane wave grows exponentially as $z \to +\infty$. If we define a
linear operator $\Lambda_0=\Lambda_0(\beta,k)$ satisfying Eq.~(\ref{dfLam}) for all $m
\ne 0$ and  
\begin{equation}
  \label{defLam0}
\Lambda_0  e^{ i\beta_0 y} = \gamma_0 e^{ i \beta_0 y},  
\end{equation}
then the resonant mode satisfies the boundary condition
\begin{equation}
  \label{bcres}
\partial_z u = \pm \Lambda_0 u, \quad z = \pm b.
\end{equation}
We emphasize that Eqs.~(\ref{uexp0}) and (\ref{bcres}) are valid for
resonant modes with only one radiation channel (i.e. the plane wave
for $m=0$) in each side of the periodic slab. 

If $\{ u, \beta, k \}$ is a resonant mode, then $\overline{u}$ (the
complex conjugate of $u$) satisfies Helmholtz equation (\ref{Eq16p})
with $k$ replaced by $\overline{k}$. If $\overline{u}$ is considered
with the original time dependence, that is, assuming $E_x = \mbox{Re}(
\overline{u}\, e^{-i \omega t})$, then it represents a solution that grows
with time. From the complex conjugate of Eq.~(\ref{uexp0}), we see
that $\overline{u}$ contains the term $e^{-i \beta y +
  \overline{\gamma}_0  z}$ for $z > b$, and it represents an
  incoming wave from $z=+\infty$. Therefore, $\overline{u}$
  represents the time-reversal of a resonant mode, it gains power
  from incoming waves and grows with time. However, $\overline{u}$ does
  not satisfy boundary condition (\ref{bcres}) with
  $\Lambda_0=\Lambda_0(\beta, \overline{k})$, because it has a
  quasi-periodicity in $y$ corresponding to $-\beta$ (instead of
  $\beta$). Nevertheless,  corresponding to the pair $(\beta, k)$ of a
  resonant mode $u$, there is another resonant mode $v$ with Bloch
  wavenumber $-\beta$ and the same $k$. The complex conjugate of $v$
  satisfies Eq.~(\ref{bcres}) with $\Lambda_0 = \Lambda_0(\beta,
  \overline{k})$. Therefore, when Eq.~(\ref{Eq16p}) is solved with 
  boundary conditions (\ref{qperiodic}) and (\ref{bcres}), we should
  obtain complex conjugate pairs $k$ and $\overline{k}$.
Assuming $\mbox{Im}(k)<0$, then the solution $u$ corresponding to $k$
is a resonant mode that decays with time, and the solution
$\overline{v}$ corresponding to $\overline{k}$ can be regarded as the
time-reversal of a resonant mode $v$ which has the same complex
frequency as $u$ but propagates in the opposite direction in $y$. 

Like the the uniform slab, the periodic slab also has improper modes
(for real $k$ and real $\beta$)  that grow exponentially as $z\to \pm
\infty$. Since the wave field can be expanded in plane waves for $|z|
> b$, there are different improper modes depending on how many of
these terms are exponentially increasing. For an improper Bloch mode
under the lightline with only one 
exponentially increasing term for $z \to +\infty$ and $z \to
-\infty$, respectively, the expansion (\ref{uexp0}) is also
valid. Therefore, these improper modes satisfy boundary condition
(\ref{bcres}). 

For the periodic slab, Eq.~(\ref{Eq16p}) also has solutions with
real $k$ (i.e. real frequency) and complex $\beta$. A leaky Bloch mode
is a solution that radiates power to the surrounding homogeneous media
(i.e. to $z = \pm \infty$) and decreases with $y$ as it propagates
forward. If a leaky mode propagates in the positive $y$ direction, then
the imaginary part of its complex Bloch wavenumber $\beta$ must be 
positive.  Assuming the leaky mode has only one radiation channel
corresponding to $m=0$, we claim that it also satisfies
Eq.~(\ref{uexp0}) and boundary condition
(\ref{bcres}). For such a leaky mode,  we have  $\mbox{Im}(\gamma_0) >  
0$, thus the plane wave $e^{i \beta y + \gamma_0 z}$ radiates power to
$z=+\infty$. Similar to the case for resonant modes, when a leaky mode
is found by solving Eq.~(\ref{Eq16p}) together with boundary
conditions (\ref{qperiodic}) and (\ref{bcres}), we also obtain a
solution with $\overline{\beta}$. If that solution is also considered
with the time dependence $e^{-i \omega t}$, it represents the
time-reversal of a leaky modes, namely, it gains power from infinity
and grows as it propagates forward.

\begin{figure*}[htb]
	\includegraphics[width=0.95\linewidth]{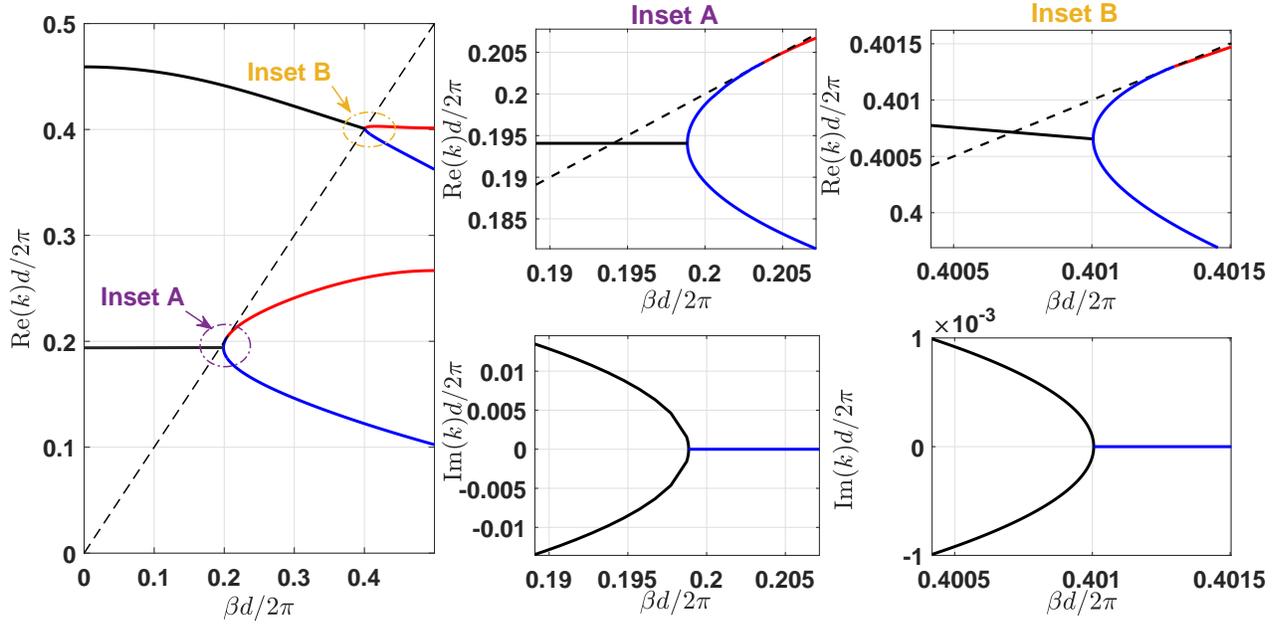}
	\caption{Dispersion curves of the first two odd modes 
          for a periodic slab with $\vare_1 = 11.56$, $\vare_2 =1$, $a 
          = 0.5\,d$ and $h = d$. The guided, resonant and improper 
          modes are shown in red, black and blue, respectively.}	
\label{Fig9}  
\end{figure*}

Using a highly accurate numerical method \cite{Amgad1}, 
we calculate guided, resonant, improper and leaky modes for a
periodic slab.  In Fig.~\ref{Fig9}, 
we show the first two odd modes (odd in $z$) for
a periodic slab with $\vare_1 = 11.56$, $\vare_2=1$,  $a = 0.5d$ and 
$h= d$. The guided, resonant and improper modes are shown in red,
black and blue, respectively. In the left panel, only the real part of
$k$ is shown for the resonant modes. The middle and right panels show
details around the starting points of the guided modes on the
lightline. It is clear that the resonant modes are not
directly connected to the guided modes. Instead, as the Bloch
wavenumber $\beta$ is increased, the resonant modes
continue their existence under the lightline (i.e., for $\mbox{Re}(k)
< |\beta|$),  end at EPs where pairs of improper modes emerge, and the
upper branches of the improper modes approach the lightline at exactly  the
same points where the guided modes emerge. Two lower panels in
Fig.~\ref{Fig9} show the imaginary parts of $k$ and
$\overline{k}$, corresponding to the resonant modes and their
time-reversals. Therefore, although the  resonant and guided modes are 
not directly connected, they are indirectly linked via EPs and improper
modes. This appears to be true for both uniform and periodic slabs. 

\begin{figure*}[htb]
	\centering  
	\includegraphics[width=0.95\linewidth]{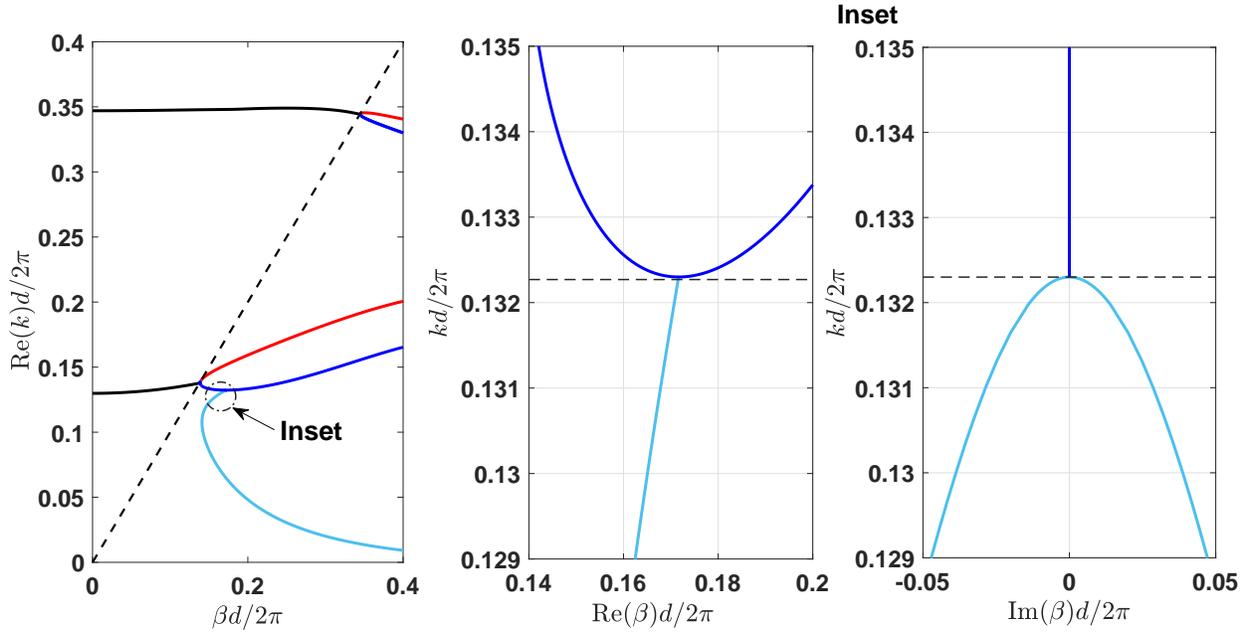}
	\caption{Dispersion curves of two even modes for a periodic  
          slab with $h = 3d$, $\vare_1 = 11.56$, $\vare_2 =1$ and $a = 
          0.5\,d$. The guided, resonant, improper and leaky modes are  
          shown in red, black, blue and cyan, respectively.}
\label{Fig10}  
\end{figure*}

From the results of Sec.~III, we expect the leaky modes to emerge from
local minima of the dispersion curves of improper modes. Since the
two improper modes shown in Fig.~\ref{Fig9} do not have local minima for
$\beta \in [0, \pi/d]$, we turn to a periodic slab with a different
parameter. In the left panel of Fig.~\ref{Fig10}, 
we show two even modes (even in $y$) for a periodic slab with thickness $h=3d$ and
the same $\vare_1$, $\vare_2$ and $a$. The red, black, blue and cyan
curves correspond to the guided, resonant, improper and leaky modes,
respectively. It is clear that the lower improper mode (with
smaller $k$) has a local minimum, and a leaky mode and its
time-reversal appear as $k$ is decreased. The minimum point is an EP
for the non-Hermitian eigenvalue problem formulated using
Eq.~(\ref{uexp0}) or Eq.~(\ref{bcres}), and it occurs at $k =
0.1323 (2\pi/d)$. The 
middle and right panels of Fig.~\ref{Fig10} show  more details around
the EP. The real and imaginary parts of $\beta$ and
$\overline{\beta}$ are shown in the middle and right panels,
respectively. As we have discussed earlier, a leaky mode is associated with
a complex $\beta$ with positive imaginary part, and its time-reversal is
associated with $\overline{\beta}$.


\section{Conclusion\label{S6}}

A uniform or periodic dielectric slab works as an optical
waveguide or a diffraction structure, depending on whether light is
propagating in or illuminated on the slab. For waveguides, it is
important to study guided modes, as well as leaky modes with complex
propagation constants. For diffraction structures, 
resonant modes are highly relevant for various applications. 
Resonant modes are usually studied above the
lightline, they approach the lightline with a nonzero imaginary part
in their complex frequencies, and are not connected with the guided modes
on the lightline. In this paper, using a more inclusive eigenvalue formulation and
highly accurate numerical methods, we found an
indirect link between the resonant and guided modes for uniform and
periodic slabs. Such a link exists below the lightline, and 
contains an EP and improper modes. We also show that the leaky modes
are related to the improper modes and emerge from EPs when frequency
is regarded as the parameter. 

Many existing studies on resonant  and
guided modes use eigenvalue formulations that are supposed
to be valid for both guided and resonant modes. Numerical studies
often use perfectly matched layers (PMLs) \cite{Berenger,PMLChew}  
that force resonant modes
(with outgoing wave components) to decay in the surrounding
homogeneous media, and be compatible with the guided
modes. Studies based on these formulations cannot find the indirect
link, because  they are incompatible with the improper modes.
We have used one formulation for the guided
modes, and another formulation for all modes that blow
up at infinity. The second formulation is valid for resonant, leaky
and improper modes, and the time-reversals of the resonant and leaky
modes, and it allows us to identify the branching points between
resonant (or leaky) and improper modes as EPs. Our study provides a
complete picture for eigenmode structures on uniform and periodic
slabs. We expect the main conclusions remain valid for more
complicated structures, including biperiodic structures such as
photonic crystal slabs.

\begin{acknowledgments}
	The second author acknowledges support from the Research Grants 
	Council of Hong Kong Special Administrative Region, China (Grant
	No. CityU 11304117).
\end{acknowledgments}

\bibliography{MyBib}

\end{document}